\documentclass[showpacs,preprintnumbers,amsmath,amssymb,prb,twocolumn]{revtex4}
\usepackage[dvips]{graphics,color}
\usepackage{dcolumn}
\begin{document}
\title{Gilbert Damping in Conducting Ferromagnets I:\\
Kohn-Sham Theory and Atomic-Scale Inhomogeneity}
\author{Ion Garate}
\author{Allan MacDonald}
\affiliation{Department of Physics, The University of Texas at Austin, Austin TX 78712}
\date{\today}

\begin{abstract}
We derive an approximate expression for the Gilbert damping coefficient $\alpha_{G}$ of itinerant electron ferromagnets which is based on their description 
in terms of spin-density-functional-theory (SDFT) and Kohn-Sham quasiparticle orbitals.
We argue for an expression in which the coupling of magnetization fluctuations 
to particle-hole transitions is weighted by the spin-dependent part of 
the theory's exchange-correlation potential, a quantity which has large spatial variations 
on an atomic length scale.  Our SDFT result for $\alpha_{G}$  
is  closely related to the previously proposed spin-torque correlation-function expression.  
\end{abstract}
\maketitle

\section{Introduction}

The Gilbert parameter $\alpha_{G}$ characterizes the damping of collective magnetization dynamics\cite{gilbert}. The key role of $\alpha_{G}$ in current-driven\cite{ralphstiles} and precessional\cite{heinrich} magnetization reversal 
has renewed interest in the microscopic physics of 
this important material parameter. It is generally accepted that in metals the damping of magnetization dynamics 
is dominated\cite{heinrich} by particle-hole pair excitation processes.  The main ideas which arise in the theory of 
Gilbert damping have been in place for some time\cite{prange,oldkambersky}.  It has however  
been difficult to apply them to real materials with the precision required for confident predictions
which would allow theory to play a larger role in designing materials with desired damping strengths. 
Progress has recently been achieved in various directions, both through studies\cite{modelstudies} of 
simple models for which the damping can be evaluated exactly and through analyses\cite{realistic} 
of transition metal ferromagnets that are based on 
realistic electronic structure calculations.  Evaluation of the torque correlation 
formula\cite{oldkambersky} for $\alpha_{G}$ used in the later calculations requires knowledge only of 
a ferromagnet's mean-field electronic structure and of its Bloch state lifetime, which makes this 
approach practical.  

Realistic {\em ab initio} theories normally
employ spin-density-functional theory\cite{sdft} which has a mean-field theory structure.
In this article we use time-dependent spin-density functional theory to 
derive an explicit expression for the Gilbert damping coefficient in terms of 
Kohn-Sham theory eigenvalues and eigenvectors.  Our final result is essentially equivalent to
the torque-correlation formula\cite{oldkambersky} for $\alpha_{G}$, but has the 
advantages that its derivation is fully consistent with density functional theory,
that it allows for a consistent microscopic treatments of both dissipative and 
reactive coefficients in the Landau-Liftshitz Gilbert (LLG) equations, and 
that it helps establish relationships between different theoretical approaches to the microscopic theory of magnetization damping.

Our paper is organized as follows.  In Section II we relate the Gilbert damping parameter $\alpha_{G}$ of a ferromagnet
to the low-frequency limit of its transverse spin response function.  Since ferromagnetism is due to 
electron-electron interactions, theories of magnetism are always many-electron theories, and it is necessary to 
evaluate the many-electron response function.  In time-dependent spin-density functional theory 
the transverse response function is calculated using a time-dependent self-
consistent-field calculation in which quasiparticles respond both to external potentials and to changes 
in the interaction-induced effective potential.  In Section III we use perturbation theory and 
time-dependent mean-field theory to express the coefficients which appear in the  
LLG equations in terms of the Kohn-Sham eigenstates and eigenvalues of the ferromagnet's ground state.
These formal expressions are valid for arbitrary spin-orbit coupling, arbitrary atomic length
scale spin-dependent and scalar potentials, and arbitrary disorder.
By treating disorder approximately, in Section IV we derive and compare two commonly used formulas for Gilbert damping. Finally, in Section V we summarize our results.

\section{Many-Body Transverse Response Function and the Gilbert Damping Parameter}

The Gilbert damping parameter $\alpha_{G}$ appears in the Landau-Liftshitz-Gilbert 
expression for the 
collective magnetization dynamics of a ferromagnet: 
\begin{equation}
\label{eq:LLG}
  \frac{\partial {\hat \Omega }}{\partial t} = {\hat \Omega} \times {\cal H}_{eff} - \alpha_{G} {\hat \Omega} \times \frac{\partial {\hat \Omega}}{\partial
  t}.
\end{equation}
In Eq.(~\ref{eq:LLG}) ${\cal H}_{eff}$ is an effective magnetic field 
which we comment on further below and $\hat{\Omega}=(\Omega_x,\Omega_y,\Omega_z)$ is the direction of the magnetization.
This equation describes the slow dynamics of smooth magnetization textures
and is formally the first term in an expansion in time-derivatives.
  
The damping parameter $\alpha_{G}$ can be measured by performing 
ferromagnetic resonance (FMR) experiments in which the magnetization 
direction is driven weakly away from an easy direction (which we take
to be the $\hat{z}$-direction.).  
To relate this phenomenological 
expression formally to microscopic theory we consider a system in 
which external magnetic fields couple only \cite{orbitalmagnetism} to the electronic 
spin degree of freedom and associate the magnetization 
direction $\hat \Omega$ with the direction of the total electron spin.
For small deviations from the easy direction, Eq.(~\ref{eq:LLG}) reads
\begin{eqnarray} 
\label{eq:linearLLG}
{\cal H}_{eff,x} &=& + \frac{\partial {\hat{\Omega}_{y}}}{\partial t} + \alpha_{G} \frac{\partial {\hat \Omega}_{x}}{\partial t} \nonumber \\ 
{\cal H}_{eff,y} &=& - \frac{\partial {\hat{\Omega}_{x}}}{\partial t} + \alpha_{G} \frac{\partial {\hat \Omega}_{y}}{\partial t} .
\end{eqnarray}  
The gyromagnetic ratio has been absorbed into the units of the field ${\cal H}_{eff}$ so that this quantity has energy units and 
we set $\hbar = 1$  throughout. 
The corresponding formal linear response theory expression is an expansion of the long wavelength transverse total spin response function
to first order \cite{FMRfrequency} in frequency $\omega$: 
\begin{equation} 
\label{eq:transverseresponse} 
S_0 {\hat \Omega}_{\alpha} = \sum_{\beta} [ \chi_{\alpha,\beta}^{st} + \omega \chi_{\alpha,\beta}^{\prime}] \; {\cal H}_{ext,\beta} 
\end{equation}
where $\alpha,\beta\in\{x,y\}$, $\omega \equiv i \partial_{t}$ is the frequency, $S_0$ is the total spin of the 
ferromagnet, ${\cal H}_{ext}$ is the external magnetic field and $\chi$ is the transverse spin-spin response function: 
\begin{widetext}
\begin{equation} 
\label{eq:Lehmann}
\chi_{\alpha,\beta}(\omega) = i \int_{0}^{\infty} dt \, \exp(i\omega t) \, \langle [S_{\alpha}(t),S_{\beta}(t)] \rangle  
= \sum_{n} \left[ \frac{\langle \Psi_0 |S_{\alpha}| \Psi_n \rangle \langle \Psi_n |S_{\beta}| \Psi_0 \rangle}{\omega_{n,0}-\omega-i\eta} 
\; + \;  \frac{\langle \Psi_0 |S_{\beta}| \Psi_n \rangle \langle \Psi_n |S_{\alpha}| \Psi_0 \rangle}{\omega_{n,0}+\omega+i\eta} \right]      
\end{equation}
Here $|\Psi_{n}\rangle$ is an exact eigenstate of the many-body Hamiltonian and 
$\omega_{n,0}$ is the excitation energy for state $n$.  We use this formal expression below to make 
some general comments about the microscopic theory of $\alpha_{G}$.
In Eq.(~\ref{eq:transverseresponse}) $\chi_{\alpha,\beta}^{st}$ is the static $(\omega=0$) limit of the response function,
and $\chi_{\alpha,\beta}^{\prime}$ is the first derivative with respect to $\omega$ evaluated at $\omega=0$.  Notice that 
we have chosen the normalization in which $\chi$ is the total spin response to a transverse field; $\chi$ is therefore extensive.
\end{widetext} 

The key step in obtaining the Landau-Liftshitz-Gilbert form for the magnetization dynamics is to recognize that in the 
static limit the transverse magnetization responds to an external magnetic field by adjusting orientation to minimize
the total energy including the internal energy $E_{int}$ and the energy due to coupling with the external magnetic field,
\begin{equation} 
E_{ext} = - S_0 {\hat \Omega} \cdot {\cal H}_{ext}.
\end{equation} 
It follows that
\begin{equation} 
\chi_{\alpha,\beta}^{st} = S_0^2 \; \left[ \frac{\partial^2 E_{int}}{\partial {\hat \Omega}_{\alpha} \, {\hat \Omega}_{\beta}}\right]^{-1}.
\end{equation}
We obtain a formal equation for $H_{eff}$ corresponding to Eq.(~\ref{eq:linearLLG}) by multiplying 
Eq.(~\ref{eq:transverseresponse}) on the left by $[\chi_{\alpha,\beta}^{st}]^{-1}$ and recognizing 
\begin{equation} 
\label{hint}
{\cal H}_{int,\alpha} =  - \frac{1}{S_0} \; \sum_{\beta} \frac{\partial^2 E_{int}}{\partial {\hat \Omega}_{\alpha} \, \partial {\hat \Omega}_{\beta}} \,\hat \Omega_{\beta} = - \frac{1}{S_0} \frac{\partial E_{int}}{\partial {\hat \Omega}_{\alpha}} 
\end{equation} 
as the internal energy contribution to the effective magnetic field ${\cal H}_{eff}={\cal H}_{int} + {\cal H}_{ext}$. 
With this identification Eq.(~\ref{eq:transverseresponse}) can be written in the form 
\begin{equation} 
H_{eff,\alpha} = \sum_{\beta} \; {\cal L}_{\alpha,\beta} \; \partial_t \hat \Omega_{\beta} 
\end{equation} 
where 
\begin{equation} 
{\cal L}_{\alpha,\beta} = -S_0 [i (\chi^{st})^{-1} \, \chi^{\prime} \, (\chi^{st})^{-1} ]_{\alpha,\beta} 
= i S_0 \partial_{\omega} \chi^{-1}_{\alpha,\beta}.
\label{eq:LLGmatrixMB}
\end{equation} 
According to the Landau-Liftshitz Gilbert equation then ${\cal L}_{x,y}=-{\cal L}_{y,x}=1$ and 
\begin{equation} 
\label{llgmanybody}
{\cal L}_{x,x} = {\cal L}_{y,y} = \alpha_{G}.
\end{equation} 
Explicit evaluation of the off-diagonal components of ${\cal L}$ will in general
yield very small deviation from the unit result assumed by the Landau-Liftshitz-Gilbert formula.
The deviation reflects mainly the fact that the magnetization magnitude varies slightly with orientation.
We do not comment further on this point because it is of little consequence.  
Similarly ${\cal L}_{x,x}$ is not in general identical to ${\cal L}_{y,y}$, although the 
difference is rarely large or important. 
Eq.(~\ref{llgmanybody}) is the starting point we use later to derive approximate expressions for $\alpha_{G}$.

In Eq.(~\ref{eq:LLGmatrixMB}) $\chi_{\alpha,\beta}(\omega)$ is the correlation function for an interacting electron 
system with arbitrary disorder and arbitrary spin-orbit coupling.  In the absence of spin-orbit coupling, but still
with arbitrary spin-independent periodic and disorder potentials, the ground state of a ferromagnet is coupled by the total spin-operator 
only to states in the same total spin multiplet.  In this case it follows from Eq.(~\ref{eq:Lehmann}) that 
\begin{equation} 
\chi^{st}_{\alpha,\beta} = 2 \sum_{n} \frac{{\rm Re}\langle \Psi_0 |S_{\alpha}| \Psi_n \rangle \langle \Psi_n |S_{\beta}| \Psi_0 \rangle]}{\omega_{n,0}} = 
\delta_{\alpha,\beta} \; \frac{S_0}{H_0} 
\end{equation} 
where $H_0$ is a static external field, which is necessary in the absence of spin-orbit coupling  
to pin the magnetization to the $\hat{z}$ direction and 
splits the ferromagnet's ground state many-body spin multiplet.  
Similarly 
\begin{equation} 
\chi^{\prime}_{\alpha,\beta} = 2 i \sum_{n}  \frac{{\rm Im}[\langle \Psi_0 |S_{\alpha}| \Psi_n \rangle \langle \Psi_n |S_{\beta}| \Psi_0 \rangle]}{\omega_{n,0}^2} = 
i \epsilon_{\alpha,\beta} \; \frac{S_0}{H_0^2}.
\end{equation} 
where $\epsilon_{x,x}=\epsilon_{y,y}=0$ and $\epsilon_{x,y}=-\epsilon_{y,x}=1$, yielding 
${\cal L}_{x,y}=-{\cal L}_{y,x}=1$ and ${\cal L}_{x,x}={\cal L}_{y,y}=0$.  Spin-orbit coupling is required for 
magnetization damping\cite{magnetic impurities}.
 
\section{SDF-Stoner Theory Expression for Gilbert Damping}

Approximate formulas for $\alpha_{G}$ in metals are inevitably based on on a self-consistent mean-field theory
(Stoner) description of the magnetic state.  Our goal is to derive an 
approximate expression for $\alpha_{G}$ when the adiabatic local spin-density approximation\cite{sdft} 
is used for the exchange correlation potential in spin-density-functional theory.
The effective Hamiltonian which describes the 
Kohn-Sham quasiparticle dynamics therefore has the form  
\begin{equation} 
{\cal H}_{KS} = {\cal H}_{P} - \Delta(n(\vec{r}),|\vec{s}(\vec{r})|) \; \hat{\Omega}(\vec{r}) \cdot \vec{s}, \; 
\end{equation} 
where ${\cal H}_{P}$ is the Kohn-Sham Hamiltonian of a paramagnetic state in which 
$|\vec{s}(\vec{r})|$(the local spin density) is set to zero, $\vec{s}$ is the spin-operator, and 
\begin{equation}
\label{delta0} 
\Delta(n,s) =  - \frac{d\; [n \epsilon_{xc}(n,s)]}{d s} 
\end{equation}
is the magnitude of the spin-dependent part of the exchange-correlation potential.
In Eq.(~\ref{delta0}) $\epsilon_{xc}(n,s)$ is the exchange-correlation energy 
per particle in a uniform electron gas with density $n$ and spin-density $s$. 
We assume that the ferromagnet is described using some semi-relativistic 
approximation to the Dirac equation like those commonly used\cite{methods} to describe 
magnetic anisotropy or XMCD, even though these approximations 
are not strictly consistent with spin-density-functional theory. Within this framework 
electrons carry only a 
two-component spin-1/2 degree of freedom and 
spin-orbit coupling terms are included in ${\cal H}_{P}$.   
Since  $n \epsilon_{xc}(n,s) \sim [(n/2+s)^{4/3} + (n/2-s)^{4/3}]$, 
 $\Delta_0(n,s) \sim n^{1/3}$ is larger closer to atomic centers and 
far from spatially uniform on atomic length scales.  This property figures prominently in 
the considerations explained below.

In SDFT the transverse spin-response function is 
expressed in terms of Kohn-Sham quasiparticle response to both 
external and induced magnetic fields: 
\begin{equation}
\label{tdmft} 
s_0(\vec{r}) \, \Omega_{\alpha}(\vec{r}) =
\int \, \frac{d \vec{r'}}{V} \; \chi^{QP}_{\alpha,\beta}(\vec{r},\vec{r'}) \; [{\cal H}_{ext,\beta}(\vec{r'}) + \Delta_{0}(\vec{r'}) \, \Omega_{\beta}(\vec{r'})].
\end{equation} 
In Eq.(~\ref{tdmft}) $V$ is the system volume, $s_0(\vec{r})$ is the magnitude of the 
ground state spin density, $\Delta_{0}(\vec{r})$ is the magnitude of the 
 spin-dependent part of the ground state exchange-correlation potential and
\begin{equation}
\chi^{QP}_{\alpha,\beta}(\vec{r},\vec{r'})=\sum_{i,j}\frac{f_{j}-f_{i}}{\omega_{i,j}-\omega-i\eta}\langle i|\vec{r}\rangle s_{\alpha} \langle\vec{r}|j\rangle \langle j|\vec{r'}\rangle s_{\beta} \langle\vec{r'}|i\rangle,
\end{equation}
where $f_{i}$ is the ground state Kohn-Sham occupation factor for eigenspinor $|i\rangle$ and $\omega_{ij} \equiv \epsilon_i-\epsilon_j$
is a Kohn-Sham eigenvalue difference. $\chi^{QP}(\vec{r},\vec{r'})$  has been normalized so that it returns the spin-density rather than total spin. 
Like the Landau-Liftshitz-Gilbert equation itself,
Eq.(~\ref{tdmft}) assumes that only the direction of the magnetization, and not the magnitudes of the charge and 
spin-densities, varies in the course of smooth collective magnetization dynamics\cite{atomic variations}.  This property should hold 
accurately as long as magnetic anisotropies and external fields are weak compared to $\Delta_0$.
We are able to use this property to avoid solving the position-space integral equation implied by 
Eq.(~\ref{tdmft}).  Multiplying by $\Delta_0(\vec{r})$ on both sides and integrating over position we find\cite{singleatom} that 
\begin{equation}
\label{chi} 
S_0 \Omega_{\alpha} = \sum_{\beta} \; \frac{1}{\bar{\Delta}_0} \; \tilde{\chi}^{QP}_{\alpha,\beta}(\omega)  \; 
\big[\Omega_{\beta} + \frac{{\cal H}_{ext,\beta}}{\bar{\Delta}_0}\big]
\end{equation}  
where we have taken advantage of the fact that in FMR experiments ${\cal H}_{ext,\beta}$ and $\hat{\Omega}$ are uniform.  $\bar{\Delta}_0$ is a spin-density weighted average of $\Delta_0(\vec{r})$,  
\begin{equation} 
{\bar \Delta}_0 = \frac{ \int d \vec{r} \Delta_0(\vec{r}) s_0(\vec{r})}{\int d \vec{r} s_0(\vec{r})},
\end{equation}
and 
\begin{equation} 
\label{chitilde}
\tilde{\chi}^{QP}_{\alpha,\beta}(\omega) = \sum_{ij} \; \frac{f_j-f_i}{\omega_{ij} - \omega - i \eta} \; 
\langle j|s_{\alpha} \Delta_0(\vec{r}) |i\rangle \, \langle i|s_{\beta} \Delta_0(\vec{r})|j\rangle
\end{equation} 
is the response function of the transverse-part of the quasiparticle exchange-correlation effective field response function,
{\em not} the transverse-part of the quasiparticle spin response function.
In Eq.(~\ref{chitilde}), $\langle i | O(\vec{r}) |j \rangle=\int d\vec{r} O(\vec{r}) \langle i|\vec{r}\rangle\langle\vec{r}|j\rangle$ denotes a single-particle matrix element.  Solving Eq.(~\ref{chi}) for the many-particle transverse susceptibility
(the ratio of $S_0 \hat{\Omega}_{\alpha}$ to $H_{ext,\beta}$) and inserting the result in Eq.(~\ref{eq:LLGmatrixMB}) yields 
\begin{equation}
\label{LLGmatrixchiqp} 
{\cal L}_{\alpha,\beta} =  i S_0 \partial_{\omega} \chi^{-1}_{\alpha,\beta} = -S_0 {\bar \Delta}_0^2 
\partial_{\omega} {\rm Im}[\tilde{\chi}^{QP\,-1}_{\alpha,\beta}].
\end{equation} 

Our derivation of the LLG equation has the advantage that the equation's reactive and dissipative 
components are considered simultaneously.  Comparing Eq.(~\ref{tdmft}) and Eq.(~\ref{hint}) we 
find that the internal anisotropy field can also be expressed in terms of $\tilde{\chi}^{QP}$:
\begin{equation} 
\label{hintchiqp}
{\cal H}_{int,\alpha} = - \bar{\Delta}_0^2 \, S_0  \; \sum_{\beta}
 \big[ \tilde{\chi}^{QP\,-1}_{\alpha,\beta}(\omega=0) - \frac{\delta_{\alpha,\beta}}{S_0 \bar{\Delta}_0} \big]\, \Omega_{\beta} .
\end{equation}
Eq.(~\ref{LLGmatrixchiqp}) and Eq.(~\ref{hintchiqp}) provide microscopic expressions for all 
ingredients that appear in the LLG equations linearized for small transverse excursions.
It is generally assumed that the damping coefficient $\alpha_{G}$ is independent of 
orientation; if so, the present derivation is sufficient.  The anisotropy-field at large transverse 
excursions normally requires additional information about magnetic anisotropy.  
We remark that if the Hamiltonian does not include a spin-dependent mean-field dipole interaction term,
as is usually the case, the above quantity will return only the magnetocrystalline anisotropy field.
Since the magnetostatic contribution to anisotropy is always well described by mean-field-theory it can be 
added separately.  

We conclude this section by demonstrating that the Stoner theory equations proposed here recover the
exact results mentioned at the end of the previous section for the limit in which spin-orbit coupling
is neglected.  We consider a SDF theory ferromagnet with arbitrary scalar and spin-dependent effective potentials.
Since the spin-dependent part of the exchange correlation potential is then the only spin-dependent term in the 
Hamiltonian it follows that 
\begin{equation} 
[{\cal H}_{KS}, s_{\alpha}] = - i \, \epsilon_{\alpha,\beta} \, \Delta_0(\vec{r}) s_{\beta} 
\end{equation} 
and hence that  
\begin{equation} 
\label{ME}
\langle i|s_{\alpha} \Delta_0(\vec{r})|j\rangle = -i \epsilon_{\alpha,\beta} \, \omega_{ij} \langle i|s_{\beta}|j\rangle .
\end{equation}
Inserting Eq.(~\ref{ME}) in one of the matrix elements of Eq.(~\ref{chitilde}) yields for the no-spin-orbit-scattering 
case 
\begin{equation} 
\label{chinoso}
\tilde{\chi}^{QP}_{\alpha,\beta}(\omega=0) = \delta_{\alpha,\beta} \; S_0 \bar{\Delta}_0.
\end{equation} 
The internal magnetic field ${\cal H}_{int,\alpha}$ is therefore 
identically zero in the absence of spin-orbit coupling and only external magnetic fields 
will yield a finite collective precession frequency.  Inserting Eq.(~\ref{ME}) in both matrix elements of 
Eq.(~\ref{chitilde}) yields
\begin{equation} 
\label{chipnoso}
\partial_{\omega} {\rm Im}[\tilde{\chi}^{QP}_{\alpha,\beta}] = \epsilon_{\alpha,\beta} S_0.
\end{equation}  
Using both Eq.(~\ref{chinoso}) and Eq.(~\ref{chipnoso}) to invert $\tilde{\chi}^{QP}$ we recover the 
results proved previously for the no-spin-orbit case using a many-body argument:
${\cal L}_{x,y}=-{\cal L}_{y,x}=1$ and ${\cal L}_{x,x}={\cal L}_{y,y}=0$.  
The Stoner-theory equations derived here allow spin-orbit interactions, and hence magnetic anisotropy and 
Gilbert damping, to be calculated consistently from the same quasiparticle response function 
$\tilde{\chi}^{QP}$.

\section{Discussion}

As long as magnetic anisotropy and external magnetic fields are weak compared to the 
exchange-correlation splitting in the ferromagnet we can use Eq.(~\ref{chinoso}) to
approximate $\tilde{\chi}^{QP}_{\alpha,\beta}(\omega=0)$.  Using this approximation 
and assuming that damping is isotropic we obtain the following explicit expression for temperature $T \to 0$:
\begin{widetext}
\begin{eqnarray}
\label{eq:alphag} 
\alpha_{G} = {\cal L}_{x,x} = -S_0 {\bar \Delta}_0^2 
\partial_{\omega} {\rm Im}[\tilde{\chi}^{QP\,-1}_{x,x}] 
&=& \frac{\pi}{S_0} \; \sum_{ij} \; \delta(\epsilon_{j}-\epsilon_F) \; 
\delta(\epsilon_i - \epsilon_{F}) \; \langle j|s_{x} \Delta_0(\vec{r}) |i\rangle \, \langle i|s_{x} \Delta_0(\vec{r})|j\rangle \nonumber \\
&=& \frac{\pi}{S_0} \; \sum_{ij} \; \delta(\epsilon_{j}-\epsilon_F) \; 
\delta(\epsilon_i - \epsilon_{F}) \; \langle j|[{\cal H}_P,s_{y}] |i\rangle \, \langle i|[{\cal H}_P,s_{y}]|j\rangle. 
\end{eqnarray} 
The second form for $\alpha_{G}$ is equivalent to the first and follows from the observation that for 
matrix elements between states that have the same energy
\begin{equation} 
\label{eq:matrixelement} 
\langle i |[{\cal H}_{KS}, s_{\alpha}]| j \rangle = - i \, \epsilon_{\alpha,\beta} \, \langle i | \Delta_0(\vec{r}) s_{\beta} 
| j \rangle + \langle i |[{\cal H}_{P}, s_{\alpha}]| j \rangle =0 \text{  (for  } \omega_{ij}=0).
\end{equation} 
Eq. (~\ref{eq:alphag}) is valid 
for any scalar and any spin-dependent potential.  It is clear however that the 
numerical value of $\alpha_{G}$ in a metal is very sensitive to the degree of disorder in 
its lattice.  To see this we observe that for a perfect crystal the Kohn-Sham eigenstates are Bloch states.
Since the operator $\Delta_0(\vec{r}) s_{\alpha}$ has the periodicity of the crystal its matrix elements
are non-zero only between states with the same Bloch wavevector label $\vec{k}$.  For the case of a perfect 
crystal then
\begin{eqnarray}
\label{eq:alphagcrystal} 
\alpha_{G} &=& \frac{\pi}{s_0} \; \int_{BZ} \frac{d\vec{k}}{(2\pi)^3} \sum_{nn'} \; \delta(\epsilon_{\vec{k}n'}-\epsilon_F) \; 
\delta(\epsilon_{\vec{k}n} - \epsilon_{F}) \; \langle \vec{k}n'|s_{x} \Delta_0(\vec{r}) |\vec{k}n\rangle \,
 \langle \vec{k}n|s_{x} \Delta_0(\vec{r})|\vec{k}n'\rangle \nonumber \\
&=& \frac{\pi}{s_0} \; \int_{BZ} \frac{d\vec{k}}{(2\pi)^3} \sum_{nn'} \; \delta(\epsilon_{\vec{k}n'}-\epsilon_F) \; 
\delta(\epsilon_{\vec{k}n} - \epsilon_{F}) \; \langle \vec{k}n'|[{\cal H}_P,s_{y}] |\vec{k}n\rangle \, 
\langle \vec{k}n|[{\cal H}_P,s_{y}]|\vec{k}n'\rangle. 
\end{eqnarray}
\end{widetext} 
where $nn'$ are band labels and $s_0$ is the ground state spin per unit volume and the integral over
$\vec{k}$ is over the Brillouin-zone (BZ). 

Clearly $\alpha_{G}$ diverges\cite{divergence} in 
a perfect crystal since $\langle \vec{k}n|s_{x} \Delta_0(\vec{r})|\vec{k}n\rangle$ is generically non-zero.
A theory of $\alpha_{G}$ must therefore always account for disorder in a crystal. 
The easiest way to account for disorder is to replace the $\delta(\epsilon_{\vec{k}n}-\epsilon_F)$ spectral
function of a Bloch state by a broadened spectral function evaluated at the Fermi energy $A_{\vec{k}n}(\epsilon_F)$.
If disorder is treated perturbatively this simple {\em ansatz} can be augmented\cite{ouralphapaper} by 
introducing impurity vertex corrections in Eq. (~\ref{eq:alphagcrystal}). Provided that the quasiparticle lifetime is computed via Fermi's golden rule, these vertex corrections restore Ward identities and yield an exact treatment of disorder in the limit of dilute impurities. Nevertheless, this approach is rarely practical outside the realm of toy models, because the sources of disorder are rarely known with 
sufficient precision.\\
Although appealing in its simplicity, the $\delta(\epsilon_{\vec{k}n}-\epsilon_F)\rightarrow A_{\vec{k}n}(\epsilon_F)$ substitution is prone to ambiguity because it gives rise to qualitatively different outcomes depending on whether it is applied to the first or second line of Eq. (~\ref{eq:alphagcrystal}):
\begin{widetext}
\begin{eqnarray}
\alpha_{G}^{(TC)}&=&\frac{\pi}{s_0} \; \int_{BZ} \frac{d\vec{k}}{(2\pi)^3} \sum_{nn'}  A_{\vec{k},n}(\epsilon_F) A_{\vec{k},n^{\prime}}(\epsilon_F) 
\langle \vec{k}n'|[{\cal H}_P,s_{y}] |\vec{k}n\rangle \, 
\langle \vec{k}n|[{\cal H}_P,s_{y}]|\vec{k}n'\rangle, 
\nonumber\\
\alpha_{G}^{(SF)}&=&\frac{\pi}{s_0} \; \int_{BZ} \frac{d\vec{k}}{(2\pi)^3} \sum_{nn'}  A_{\vec{k},n}(\epsilon_F) A_{\vec{k},n^{\prime}}(\epsilon_F) 
\langle \vec{k}n'|s_{x} \Delta_0(\vec{r}) |\vec{k}n\rangle \,
 \langle \vec{k}n|s_{x} \Delta_0(\vec{r})|\vec{k}n'\rangle. \nonumber \\
\end{eqnarray}
\end{widetext}
$\alpha_{G}^{(TC)}$ is the torque-correlation (TC) formula used in realistic electronic structure calculations\cite{realistic} and $\alpha_{G}^{(SF)}$ is the spin-flip (SF) formula used in certain toy model calculations\cite{jairo}. The discrepancy between TC and SF expressions stems from inter-band ($n\neq n^{\prime}$) contributions to damping, which may now connect states with \emph{different} band 
energies due to the disorder broadening of the spectral functions. Therefore, $\langle \vec{k} n |[{\cal H}_{KS}, s_{\alpha}]| \vec{k} n^{\prime} \rangle$ no longer vanishes for $n\neq n^{\prime}$ and  Eq. (~\ref{eq:matrixelement}) indicates that $\alpha_{G}^{(TC)}\simeq\alpha_{G}^{(SF)}$ only if the Gilbert damping is dominated by intra-band contributions and/or if the energy difference between the states connected by inter-band transitions is small compared to $\Delta_{0}$. When $\alpha_{G}^{(TC)}\neq\alpha_{G}^{(SF)}$, it is \emph{a priori} unclear which approach is the most accurate. One obvious flaw of the SF formula is that it produces a spurious damping in absence of spin-orbit interactions; this unphysical contribution originates from inter-band transitions and may be cancelled out by adding the leading order impurity vertex correction\cite{tatara}. In contrast, $[{\cal H}_{P},s_{y}]=0$ in absence of spin-orbit interaction and hence the TC formula vanishes identically, even without vertex corrections. From this analysis, TC appears to have a pragmatic edge over SF in materials with weak spin-orbit interaction. However, insofar as it allows inter-band transitions that connect states with $\omega_{i,j}>\Delta_{0}$, TC is not quantitatively reliable. Furthermore, it can be shown\cite{ouralphapaper} that when the intrinsic spin-orbit coupling is significant (e.g. in ferromagnetic semiconductors), 
the advantage of TC over SF (or vice versa) is marginal, and impurity vertex corrections play a significant role.  
\bigskip

\section{Conclusions}

Using spin-density functional theory we have derived a Stoner model expression for the Gilbert damping coefficient in itinerant ferromagnets. This expression accounts for atomic scale variations of the exchange self energy, as well as for arbitrary disorder and spin-orbit interaction. By treating disorder approximately, we have derived the spin-flip and torque-correlation formulas previously used in toy-model and \emph{ab-initio} calculations, respectively.  We have traced the discrepancy between these equations to the treatment of inter-band transitions that connect states which are not close in energy. A better treatment of disorder, which requires the inclusion of impurity vertex corrections, will be the ultimate judge on the 
relative reliability of either approach.  When damping is dominated by intra-band transitions, a circumstance which we believe 
is common, the two formulas are identical and both are likely to provide reliable estimates.
This work was suported by the National Science Foundation under grant DMR-0547875.


\begin{thebibliography}{20}
\bibitem{gilbert} For a historical perspective see T.L. Gilbert, IEEE Trans. Magn. \textbf{40}, 3443 (2004).
\bibitem{ralphstiles} For an introductory review see D.C. Ralph and M.D. Stiles, 
J. Magn. Mag. Mater. {\bf 320}, 1190 (2008).
\bibitem{heinrich} J.A.C. Bland and B. Heinrich (Eds.), {\em Ultrathin Magnetic Structures III: Fundamentals of
Nanomagnetism} (Springer-Verlag, New York, 2005).
\bibitem{prange} V. Korenman and R. E. Prange, Phys. Rev. B {\bf 6}, 2769 (1972).
\bibitem{oldkambersky} V. Kambersky, Czech J. Phys. B \textbf{26}, 1366 (1976).
\bibitem{modelstudies} Y. Tserkovnyak, G.A. Fiete, and B.I. Halperin, Appl. Phys. Lett. {\bf 84},
5234 (2004); E.M. Hankiewicz, G. Vignale and Y. Tserkovnyak, Phys. Rev. B \textbf{75}, 174434 (2007); 
Y. Tserkovnyak \emph{et al.}, Phys. Rev. B \textbf{74}, 144405 (2006) ; 
H.J. Skadsem, Y. Tserkovnyak, A. Brataas, G.E.W. Bauer, Phys.
Rev. B {\bf 75}, 094416 (2007);
H. Kohno, G. Tatara and J. Shibata, J. Phys. Soc. Japan \textbf{75}, 113706 (2006); R.A. Duine \emph{et al.}, Phys. Rev. B \textbf{75}, 214420 (2007).
Y. Tserkovnyak, A. Brataas, and G.E.W. Bauer, J. Magn. Mag. Mater. {\bf 320}, 1282 (2008).
\bibitem{realistic}  K. Gilmore, Y.U. Idzerda and M.D. Stiles, Phys. Rev. Lett. \textbf{99}, 27204 (2007);
 V. Kambersky, Phys. Rev. B \textbf{76}, 134416 (2007).
\bibitem{magnetic impurities} For zero spin-orbit coupling $\alpha_{G}$ vanishes even in presence of magnetic impurities, provided that their spins follow the dynamics of the magnetization adiabatically. 
\bibitem{sdft} O. Gunnarsson, J. Phys. F {\bf 6}, 587 (1976). 
\bibitem{vignalesdftdynamics} Z. Qian, G. Vignale, Phys. Rev. Lett. {\bf 88}, 056404 (2002).
\bibitem{orbitalmagnetism} In doing so we dodge the subtle difficulties which complicate theories of 
orbital magnetism in bulk metals.  See for example J. Shi, G. Vignale, D. Xiao, and Q. Niu,
Phys. Rev. Lett. {\bf 99}, 197202 (2007); I. Souza and D. Vanderbilt, 
Phys. Rev. B {\bf 77}, 054438 (2008) and work cited therein.  This simplification should 
have little influence on the theory of damping because the orbital contribution to the 
magnetization is relatively small in systems of interest and because it in any event 
tends to be collinear with the spin magnetization. 
\bibitem{FMRfrequency} For most materials the FMR frequency is by far the smallest energy scale in the problem.
Expansion to linear order is almost always appropriate. 
\bibitem{methods} See for example A.C. Jenkins and W.M. Temmerman,
Phys. Rev. B {\bf 60}, 10233 (1999) and work cited therein.
\bibitem{atomic variations} This approximation does not preclude strong spatial variations of $|s_{0}(\vec{r})|$ and $|\Delta_{0}(\vec{r})|$ at atomic lenghtscales; rather it is assumed that such spatial profiles will remain unchanged in the course of the magnetization dynamics.
\bibitem{singleatom}  For notational simplicity we assume that all magnetic atoms are identical.  Generalizations to 
magnetic compounds are straight forward.
\bibitem{divergence} Eq. (~\ref{eq:alphag}) is valid provided that $\omega\tau<<1$. While this condition is normally satisfied in cases of practical interest, it invariably breaks down as $\tau\rightarrow\infty$. Hence the divergence of  Eq. (~\ref{eq:alphag}) in \emph{perfect} crystals is spurious. 
\bibitem{ouralphapaper} I. Garate and A.H. MacDonald (in preparation).
\bibitem{jairo} J. Sinova \emph{et al.}, Phys. Rev. B \textbf{69}, 85209 (2004). In order to get the equivalence, trade $h_{z}$ by $\Delta_{0}$ and use $\Delta_{0}=J_{pd} S_{0}$, where $J_{pd}$ is the p-d exchange coupling between GaAs valence band holes and Mn d-orbitals. In addition, note that our spectral function differs from theirs by a factor $2\pi$.   
\bibitem{tatara} H. Kohno, G. Tatara and J. Shibata, J. Phys. Soc. Japan \textbf{75}, 113706 (2006).


\end{thebibliography}
\end{document}